\documentclass[aps,prl,twocolumn,longbibliography,superscriptaddress,floatfix]{revtex4-2}

\usepackage{amsmath}
\usepackage{graphicx}
\usepackage{mathtools}
\usepackage{xcolor}
\usepackage{float}
\usepackage{placeins}
\usepackage{bm}
\usepackage[normalem]{ulem}
\usepackage{dcolumn}
\usepackage{xr}

\graphicspath{{figs/}}
\usepackage[percent]{overpic}

\begin{document}

\title{Frustration-Assisted Electrical Switching in Co$_{x}$TaS$_2$}

\newcommand{\BGU}{Department of Physics, Ben-Gurion University of the Negev, Be'er-Sheva 84105, Israel}
\newcommand{\NRCN}{Physics Department, NRCN, P.O. Box 9001, Be'er Sheva, 84190, Israel}

\newcommand{\NHMFL}{National High Magnetic Field Laboratory, Tallahassee, Florida 32310, USA} 

\author{Matan Sterenberg}
\affiliation{\BGU}

\author{Bar Segura}
\affiliation{\BGU}

\author{Muntaser Naamneh}
\affiliation{\BGU}

\author{Eran Maniv}
\affiliation{\BGU}


\date{\today}

\begin{abstract}

Electrical switching in antiferromagnets is commonly described as the action of current-induced torques on a rigid order parameter. In frustrated magnets, however, competing states can create a history-dependent energy landscape in which slow collective dynamics actively participate in electrical reconfiguration. Here, we demonstrate frustration-assisted switching in near-stoichiometric Co$_x$TaS$_2$, a van der Waals antiferromagnet located at the composition-sensitive boundary between Hall-active and Hall-inactive multi-Q-related magnetic regimes. Although the crystals exhibit no resolved spontaneous Hall response, they retain van Hove singularities near the Fermi level and order antiferromagnetically below T$_N$ $\approx$ 31.5 K. Current pulses write nonvolatile anisotropic resistance states in microstructured devices, with an activation current density nearly an order of magnitude below that reported for the Hall-active lower-intercalation phase. The switching response separates into two temperature-dependent dynamical regimes rather than following a single current threshold. Independent magnetization-aging and post-field-relaxation measurements reveal slow, history-dependent magnetic dynamics, and the relaxation strength closely follows the inverse switching amplitude. These results identify magnetic frustration and glassy dynamics as central ingredients in the electrical reconfiguration of Co$_x$TaS$_2$, establishing a route to functional switching through a dynamically soft antiferromagnetic landscape.

\end{abstract}

\maketitle

\section{INTRODUCTION}

Electrical control of antiferromagnetic (AFM) order is typically formulated as a torque problem: a current-induced spin polarization acts on a well-defined Néel vector, while the resulting state is detected through anisotropic magnetoresistance or Hall transport~\cite{Baltz2018RMP}. This picture has enabled landmark demonstrations in CuMnAs, Mn$_2$Au and related systems~\cite{Wadley2016Science,bodnar2018writing,Nair2020NatMater}, but it implicitly treats the magnetic order as comparatively rigid. In frustrated magnets, the relevant degrees of freedom can instead include nearly degenerate textures, metastable domains, and slow collective modes~\cite{mezard1984nature,binder1986spin}. These features are usually regarded as obstacles to deterministic switching, yet they may also reduce the barriers separating electrically distinguishable states~\cite{FeNbS2SpinGlassSciAdv, tserkovnyak2021collective}. Establishing whether magnetic frustration can actively assist current-driven reconfiguration is therefore both a fundamental question and a materials-design opportunity for AFM spintronics.

Magnetically intercalated transition-metal dichalcogenides (ITMDs) provide a particularly revealing setting for this problem because small changes in intercalant concentration can reorganize the magnetic ground state, its domain structure, and its dynamical response. Fe$_{1/3}$NbS$_2$ first established that an AFM ITMD can be electrically written and read through an anisotropic-magnetoresistance-like (AMR-like) response at unusually low current densities~\cite{Nair2020NatMater}. Subsequent work showed that this functionality cannot be understood solely as the rotation of a rigid Néel vector: switching is strongest away from exact one-third filling, where AFM order coexists with a glass state, exchange bias, slow relaxation, and collective nonlocal reconfiguration~\cite{FeNbS2SpinGlassSciAdv,EM,haley2023long}. These results identified frustrated, history-dependent degrees of freedom as active participants in electrical switching. More recent studies have broadened the range of control mechanisms available in ITMDs, including current-driven manipulation of helical and chiral textures, carrier-density tuning of topological multi-Q order, and Joule-heating-assisted magnetic writing~\cite{NiNbS2HelicalPRL2025,xiong2025all,zhang2026current, Rodriguez2026HAMR}. Together, these developments show that pulsed-current control in ITMDs is not governed by a universal torque mechanism; instead, it depends on how the current couples to the specific magnetic energy landscape.

Co$_x$TaS$_2$ offers a unique controlled platform in which to test this idea because its magnetic order changes sharply across a very narrow composition range near $x=0.33$. Slightly Co-deficient crystals host two magnetic transitions and a large spontaneous Hall response associated with non-coplanar multi-Q order and finite scalar-spin-chirality, whereas compositions closer to and above one-third suppress the net zero-field Hall signal~\cite{takagi2023spontaneous,CoTaS2TripleQNatCommun2023,Park2024CoTaS2Composition,Noh2026CoTaS2OrderingARPES}. This Hall-inactive regime was initially assigned to a coplanar helical state with Q = (1/3,0,0)~\cite{Park2024CoTaS2Composition}, but recent neutron and optical studies show that the distinction is subtler: overdoped Co$_{0.336}$TaS$_2$ can retain a 3Q configuration with spatially alternating scalar chirality, so that local chiral textures survive while the macroscopic Hall response cancels~\cite{Cho2026CoTaS2AlternatingChirality3Q}. Complementary optical and transport studies further reveal intertwined chiral, nematic, and anisotropic multi-Q states in closely related compositions, underscoring how weak changes in filling or intercalant order reorganize the balance between competing magnetic configurations~\cite{kirstein2026tunable,Kruppe2026AnisotropicMultiQCoTaS2,feng2025nonvolatile,hu2026tunable}. Co$_x$TaS$_2$ therefore provides a rare opportunity to compare electrical switching across closely related magnetic states within the same host lattice, while largely avoiding the ambiguities inherent to comparisons between chemically distinct materials.

\begin{figure*}
    \centering
    \includegraphics[width=\textwidth]{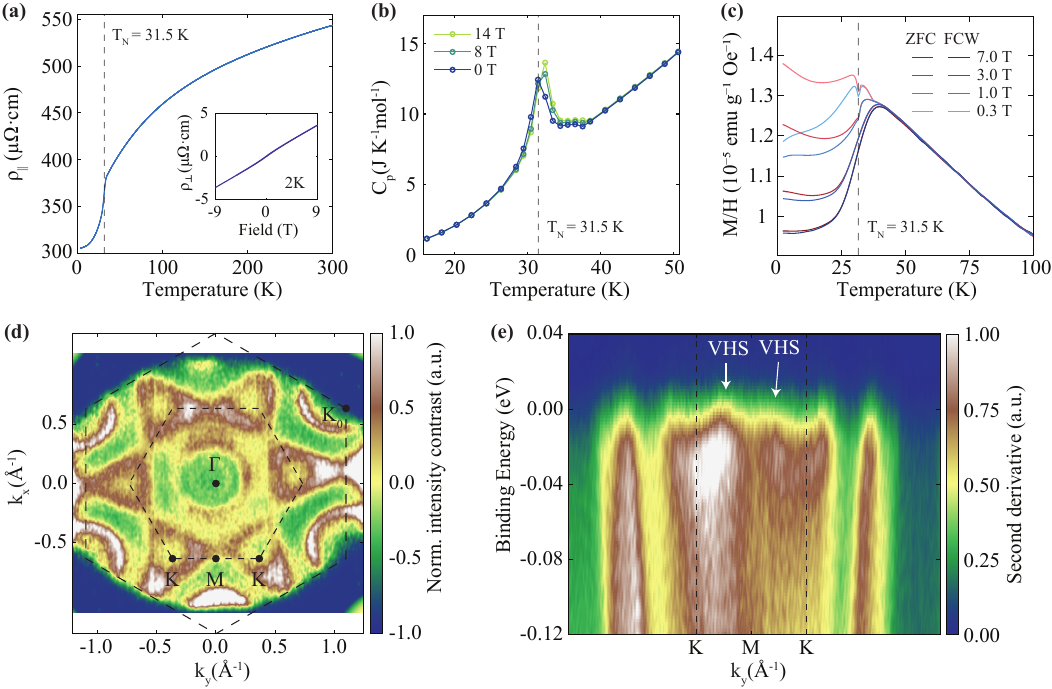}
    \caption{
    \textbf{Characterization of near-stoichiometric Hall-inactive Co$_x$TaS$_2$ near $x=0.33$.}
    \textbf{(a)} Longitudinal resistivity $\rho_{\parallel}(T)$ measured on a microdevice from 2--300~K. The dashed vertical line marks the AFM transition temperature $T_N=31.5$~K. Inset: transverse resistivity $\rho_{\perp}$ versus out-of-plane magnetic field, measured at 2~K between $-9$ and $9$~T on an additional cross-shaped switching microdevice; no resolved zero-field spontaneous Hall component is observed.
    \textbf{(b)} Heat capacity $C_p(T)$ of the parent bulk crystal measured in out-of-plane magnetic fields of 0, 8, and 14~T. The sharp anomaly at $T_N=31.5$~K defines the magnetic transition used throughout the paper.
    \textbf{(c)} Magnetic susceptibility $M/H$ measured under zero-field-cooled (ZFC) and field-cooled-warming (FCW) protocols for applied fields of 0.3, 1.0, 3.0, and 7.0~T. The dashed vertical line marks $T_N=31.5$~K.
    \textbf{(d)} ARPES constant-energy intensity map near the Fermi level, measured at 7~K using 80~eV photons with LH polarization and shown as normalized intensity contrast in the $(k_x,k_y)$ plane. Dashed lines indicate the hexagonal Brillouin-zone boundary and high-symmetry points.
    \textbf{(e)} Second-derivative ARPES intensity along the $K$--$M$--$K$ cut under the same measurement conditions as panel~(d). Arrows mark two van-Hove-related features near the Fermi level.
    }
    \label{fig:combined}
\end{figure*}

Here we investigate Co$_x$TaS$_2$ with $x \approx 0.33$, a near-stoichiometric composition positioned at this sharply tuned magnetic boundary. Single-crystal X-ray diffraction (SCXRD) confirms the expected Co-intercalated structure, while energy-dispersive X-ray spectroscopy (EDX) indicates a composition close to $x \approx 0.33$ (see Supplementary Information, Sample growth and composition, and Tables~S1--S2). Transport shows no resolved spontaneous Hall component (Fig.~\ref{fig:combined}a inset), placing the sample on the Hall-inactive side of the phase diagram. At the same time, angle-resolved photoemission spectroscopy (ARPES) reveals an inverse-Mexican-hat-like dispersion and two van Hove singularities within a few meV of the Fermi level, retaining the electronic fingerprints associated with the nearby multi-Q instability~\cite{Luo2026VanHoveCoTaS2}. We find that current pulses nevertheless write nonvolatile anisotropic resistance states, with switching activated at $J \approx (2\text{--}3)\times10^{5}~\mathrm{A\,cm^{-2}}$, nearly an order of magnitude below the activation scale reported for the Hall-active lower-intercalation phase~\cite{xiong2025all,zhang2026current}. More importantly, the response separates into two distinct temperature-dependent dynamical regimes rather than following a single current threshold. Magnetization aging and post-field relaxation independently reveal slow, history-dependent dynamics whose temperature evolution correlates with the switching response. These results show that electrical reconfiguration persists after the net Hall response is suppressed and support a picture in which frustration and glassy magnetic dynamics lower the barriers between current-written states.

\section{RESULTS and DISCUSSION}

We first establish the magnetic and electronic character of the crystals used for electrical switching. The longitudinal resistivity exhibits a clear anomaly near 31.5~K (Fig.~\ref{fig:combined}a), coincident with the sharp heat-capacity feature that defines the AFM transition temperature T$_N$ (Fig.~\ref{fig:combined}b). The susceptibility shows a corresponding cusp together with a field-dependent separation between zero-field-cooled and field-cooled-warming traces below T$_N$ (Fig.~\ref{fig:combined}c). The lowest-field susceptibility also exhibits a pronounced nonmonotonic anomaly similar to that reported near the composition-driven boundary between competing antiferromagnetic phases in Co$_x$TaS$_2$~\cite{Park2024CoTaS2Composition}. The ZFC--FCW separation is strongest at low field and progressively suppressed on increasing field, possibly suggesting that long-range AFM order coexists with a slowly relaxing or frozen magnetic component. By contrast, the low-temperature transverse resistivity remains non-hysteretic within our experimental resolution, with no resolved spontaneous Hall contribution (inset of Fig.~\ref{fig:combined}a). The present composition therefore lies outside the Hall-active regime established for more Co-deficient Co$_x$TaS$_2$, despite retaining a well-defined ordered state below T$_N$~\cite{Park2024CoTaS2Composition,CoTaS2TripleQNatCommun2023}.

The electronic structure reveals why this Hall-inactive state remains close to magnetic instability. ARPES resolves an inverse-Mexican-hat-like dispersion along $K$–$M$–$K$, producing two nearby van Hove singularities and a band extremum within $\sim 10$ meV of the Fermi level (Fig.~\ref{fig:combined}d,e). Such a high density of low-energy electronic states strongly enhances the sensitivity of the magnetic ground state to small changes in filling and intercalant occupancy, and has been identified as an electronic fingerprint of the multi-Q instability in Co$_x$TaS$_2$~\cite{Luo2026VanHoveCoTaS2}. The coexistence of a sharp AFM transition, low-field magnetic history dependence, a suppressed spontaneous Hall response, and van Hove singularities near $E_F$ places Co$_x$TaS$_2$ near $x=0.33$ at a compositionally frustrated boundary: macroscopically Hall-inactive, yet electronically proximate to the competing multi-Q-related state found at slightly lower intercalation.

\begin{figure*}
    \centering
    \includegraphics[width=\textwidth]{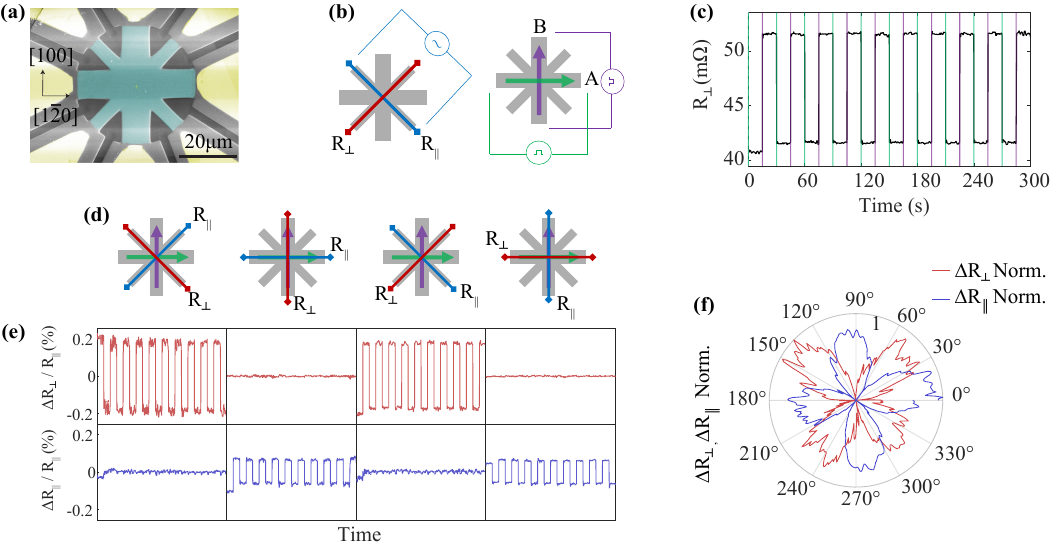}
\caption{
    \textbf{Electrical switching in Hall-inactive Co$_x$TaS$_2$ near $x=0.33$.}
    \textbf{(a)} SEM image of the cross-shaped focused-ion-beam microdevice. The horizontal and vertical arms define the fixed write-pulse directions A and B, while the diagonal arms provide the main readout paths.
    \textbf{(b)} Write/read schematic for the diagonal-readout configuration used in panel~(c). Alternating unipolar write pulses are applied along A and B. The resistance state is read using a 0.1~mA AC probe current along one diagonal bar. The longitudinal and transverse readout channels are parallel and perpendicular to the AC probe current, respectively. The probe-current direction is rotated by 45 degrees relative to the A/B write-pulse axes. Unless stated otherwise, this same diagonal-readout geometry is used for the switching traces and amplitudes reported throughout the main text.
    \textbf{(c)} Representative transverse-resistance trace measured at 4~K using 35~mA, 10~ms alternating A/B write pulses with a 15~s wait after each pulse. Ten pulses were applied along each write direction, giving 20 total write pulses. The trace switches reproducibly between two nonvolatile resistance states.
    \textbf{(d)} Readout-geometry dependence of the switching response. The write-pulse directions remain fixed while the readout configuration is rotated, transferring the switching contrast between the transverse and longitudinal channels and reversing its sign for symmetry-related readout orientations.
    \textbf{(e)} Normalized switching amplitudes extracted from the transverse and longitudinal readout channels for the readout geometries shown in panel~(d), measured at 4~K using the same 35~mA, 10~ms alternating-pulse protocol as in panel~(c).
    \textbf{(f)} In-plane magnetic-field rotation dependence of the Hall-bar magnetoresistance, measured at 4~K and 11~T. Both longitudinal and transverse components are normalized for comparison. The plotted transverse component is obtained after subtracting a fitted sinusoidal background, providing an independent reference for the anisotropic transport symmetry. Positive values of $\Delta R_{\perp}/R_{\parallel}$ follow the transverse switching sign convention used throughout the main text. Raw longitudinal and transverse data and full analysis are presented in Supplementary Fig.~S3.
}
    \label{fig:switching_overview}
\end{figure*}

We next test whether the Hall-inactive ordered state can be electrically reconfigured. A cross-shaped microdevice was patterned from the same crystal growth, allowing current pulses to be applied along two fixed in-plane directions while the written state is read independently through longitudinal and transverse resistance channels (Fig.~\ref{fig:switching_overview}a,b). Low-amplitude AC excitation is used only for readout ($\sim$100~$\mu$A), whereas 35~mA write pulses are applied sequentially along the two orthogonal arms of the device. The principal switching observable is the nonvolatile transverse resistance change normalized by the longitudinal resistance, $\Delta R_{\perp}/R_{\parallel}$. Further experimental details are provided in the Supporting Information, including FIB device geometry (Fig.~S2), the switching protocol, and the current-density estimate.

Alternating the write direction reproducibly drives the device between two resistance states (Fig.~\ref{fig:switching_overview}c). The resistance is measured only after the pulse and after the immediate transient has decayed, demonstrating that the signal is retained in the absence of the writing current. Repeated pulses along the same direction reinforce the same state, whereas changing the writing direction reverses the resistance contrast. The effect is observed in both longitudinal and transverse channels and is reproduced on an additional device, with the expected transfer between readout channels across multiple configurations (see Supplementary Fig.~S5). The switching effect is connected to the AFM phase via the temperature and magnetic field dependencies; the effect vanishes above the AFM transition temperature (Supplementary Figs. S6 and S7) and persists stably at high magnetic fields up to 14 T (Supplementary Figs. S10 and S11). The onset of switching occurs at current densities of approximately
$(2\text{--}3)\times10^{5}~\mathrm{A\,cm^{-2}}$, with larger pulse amplitudes and widths producing a substantially enhanced response (see Supplementary Figs. S8 and S9). 

\begin{figure*}
    \centering
    \includegraphics{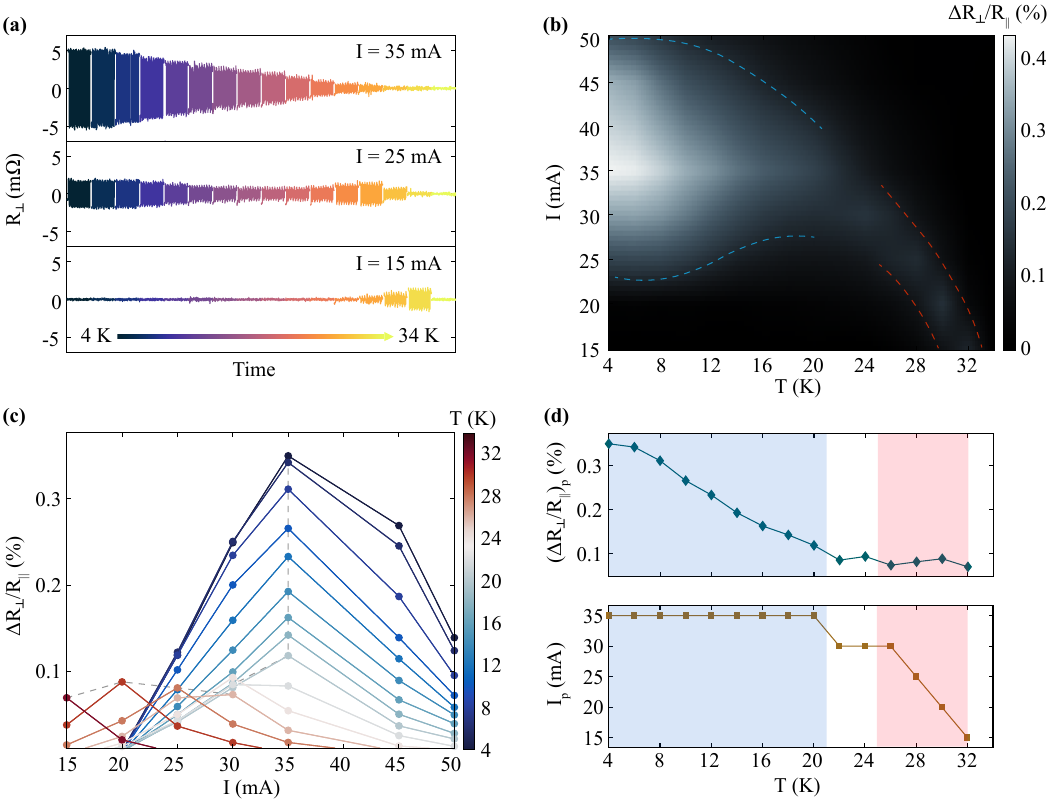}
     \caption{
    \textbf{Temperature- and current-dependent switching reveals two dynamical regimes.}
    \textbf{(a)} Representative transverse-resistance traces $R_{\perp}(t)$ measured on the cross-shaped switching microdevice for write-pulse currents of 35, 25, and 15~mA. Traces are colored by temperature from 4 to 34~K. The traces illustrate two response lobes: a stronger low-temperature/high-current lobe and a weaker high-temperature/low-current lobe, with an intermediate temperature range in which both contributions are visible.
    \textbf{(b)} Switching amplitude $\Delta R_{\perp}/R_{\parallel}$, expressed as a normalized peak-to-peak percentage, as a function of temperature $T$ and write-pulse current $I$. Normalized peak-to-peak switching amplitudes are reported as positive magnitudes. The displayed map is interpolated from the measured post-pulse switching amplitudes to visualize the ridge-like response in the current--temperature plane; the corresponding measured-grid map and peak-extraction robustness are shown in Supplementary Fig.~S13.
    \textbf{(c)} Constant-temperature current cuts of the switching response at all measured temperatures, plotted directly from the measured switching amplitudes before interpolation. The response is largest at low temperature near $I\simeq35$~mA and weakens and shifts to lower current near $T_N$.
    \textbf{(d)} Peak switching amplitude $(\Delta R_{\perp}/R_{\parallel})_{p}$ and corresponding peak current $I_{p}$ as functions of temperature. For each measured temperature, $(\Delta R_{\perp}/R_{\parallel})_{p}$ is the maximum measured switching amplitude over write-pulse current, and $I_{p}$ is the current at which that maximum occurs. The blue shaded range marks the low-temperature region where the ridge in panel~(b) is strongest and comparatively stable, while the red shaded range marks the softer near-transition switching region associated with the low-current contour of the map.
    }
    \label{fig:switching_map}
\end{figure*}

\begin{figure*}[!t]
    \centering
    \includegraphics[width=\textwidth]{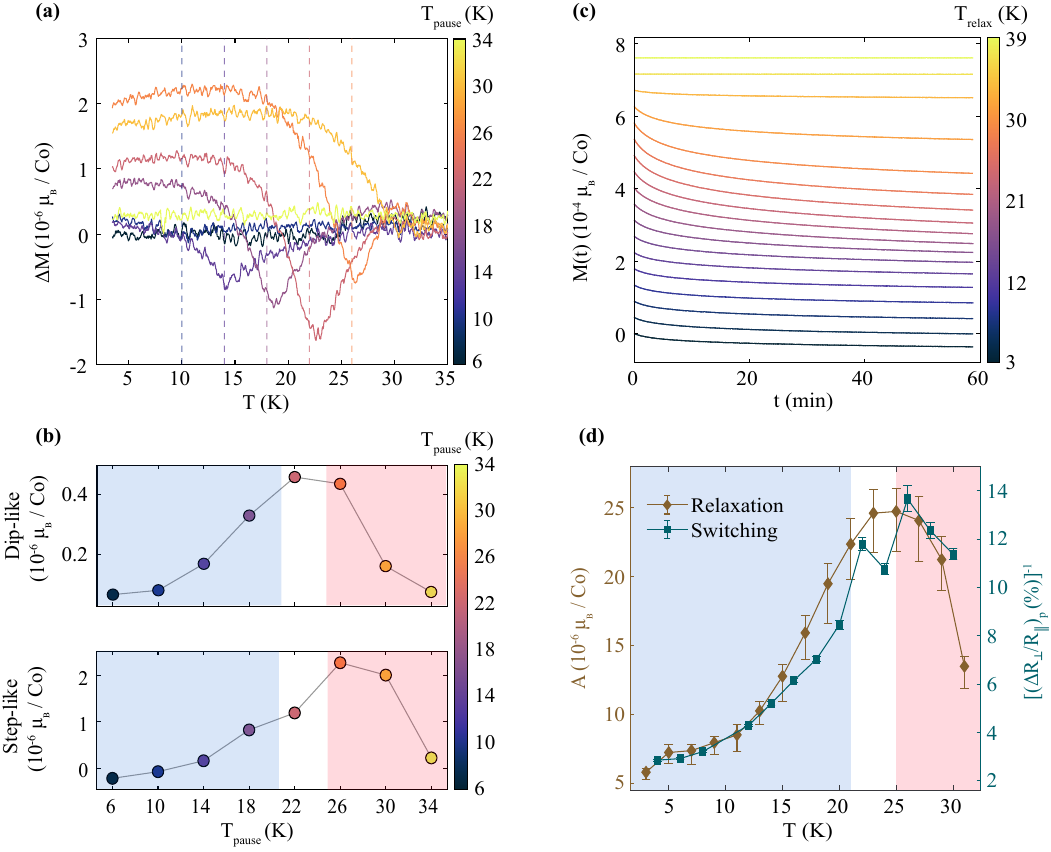}
    \caption{
    \textbf{Magnetic aging, post-field relaxation, and comparison to electrical switching.}
    \textbf{(a)} Aging response $\Delta M(T)$ obtained by subtracting the no-pause zero-field-cooled warming reference from zero-field-cooled warming curves measured after a 60~min pause at $T_{\mathrm{pause}}$,
    $\Delta M(T)=M_{\mathrm{ZFC}}^{\mathrm{pause}}(T)-M_{\mathrm{ZFC}}^{\mathrm{no\ pause}}(T)$. Raw data are presented in Supplementary Fig. S14.
    Curves are colored by $T_{\mathrm{pause}}=6$, 10, 14, 18, 22, 26, 30, and 34~K; dashed vertical lines mark the corresponding pause temperatures.
    \textbf{(b)} Phenomenological aging components extracted from panel~(a) as a function of $T_{\mathrm{pause}}$: Dip-like memory amplitude (top) and Step-like amplitude (bottom). The labels denote line-shape components of the aging signal and are not assigned to distinct microscopic phases.
    \textbf{(c)} Post-field-off relaxation traces $M(t)$ measured after field cooling in 1~T along the crystallographic $c$ axis, then removing the field and recording the magnetization for 60~min at fixed relaxation temperatures $T_{\mathrm{relax}}=3$--39~K. Curves are vertically offset for clarity.
    \textbf{(d)} Empirical relaxation strength $A(T)$ compared with the inverse peak switching amplitude $[(\Delta R_{\perp}/R_{\parallel})_{p}]^{-1}$, with the switching amplitude expressed in $\%^{-1}$. The two quantities closely track one another over the ordered-state temperature range. The relaxation strength is quantified as $A(T)=\left[\mathrm{median}_t\,\Delta M(T,t)^2\right]^{1/2}$ over the post-field-off measurement window, where $\Delta M(T,t)$ is referenced to the late-time magnetization near the end of the 60~min trace. Full analysis details are provided in the Supporting Information, including the switching peak-extraction robustness (Supplementary Fig.~S13), relaxation protocol, and relaxation-amplitude robustness (Supplementary Fig.~S15).
    }
    \label{fig:afm_fm}
\end{figure*}

The relation between the written state and the readout geometry is resolved by rotating the AC probe configuration while keeping the two writing directions fixed (Fig.~\ref{fig:switching_overview}d,e). As the angle between the writing pulse and probe current is varied, the switching signal transfers continuously between the transverse and longitudinal channels and reverses sign over a $90^{\circ}$ rotation. The two channels therefore represent different projections of the same anisotropic resistivity tensor rather than independent switching effects. This geometry dependence closely resembles the AMR readout established in CuMnAs, Fe$_x$NbS$_2$, and other electrically switchable antiferromagnets, where a written redistribution of magnetic domains rotates the principal axes of the resistivity tensor~\cite{Wadley2016Science, Nair2020NatMater, OriginsAMRFeNbS2PRB2021}.

A direct comparison is provided by the magnetic-field-written AMR in Fig.~\ref{fig:switching_overview}f. Rotating an in-plane field relative to the low-current probe produces the same transfer of spectral weight between R$_{\parallel}$ and $R_{\perp}$, including the corresponding sign reversal. The agreement shows that electrical and magnetic writing are detected through the same anisotropic transport channel. Importantly, this correspondence identifies the symmetry of the written state but does not by itself determine the microscopic current-induced torque. In the present composition, where the magnetic structure may involve a compensated multi-Q texture rather than a single Néel vector, the pulses may redistribute domains, alter the balance among nearly degenerate textures, or couple to slower collective degrees of freedom. The temperature and current dependence examined below provides the more direct evidence that the writing process is controlled by such a non-rigid magnetic landscape.

Having established nonvolatile electrical writing, we next examine how the response evolves across the ordered phase. Figure~\ref{fig:switching_map} maps the switching amplitude as a function of temperature and pulse current and reveals that the response cannot be described by a single activation threshold. Instead, the maximum switching follows a pronounced ridge in the current–temperature plane, separating a robust low-temperature regime from a distinct response that develops on approaching T$_N$.

The representative traces in Fig.~\ref{fig:switching_map}a already expose this crossover. At 35~mA, the resistance contrast is largest at low temperature and steadily weakens on warming. At 15 mA, by contrast, the response is nearly absent deep in the ordered phase and emerges only at elevated temperature. Intermediate currents bridge these limits. Thus, temperature does not simply suppress an otherwise fixed switching process. It changes the range of current over which the magnetic state can be reconfigured, indicating that the barriers and accessible configurations of the underlying magnetic landscape evolve substantially within the AFM phase.

This evolution is quantified by the switching map in Fig.~\ref{fig:switching_map}b (full raw traces are presented in Supplementary Fig.~S12, and the corresponding measured-grid map and peak-extraction robustness are shown in Supplementary Fig.~S13). At low temperature, the response is concentrated around a comparatively high and nearly temperature-independent current scale, where the switching amplitude reaches its maximum (blue region). On warming, the ridge moves toward lower pulse current while its amplitude decreases (red region). The nonmonotonic dependence on current is equally significant: beyond the ridge, increasing the drive does not produce a larger written state and can instead reduce the nonvolatile contrast. The current therefore acts neither as a simple linear control parameter nor as a fixed thermal threshold. Rather, its effectiveness depends on the magnetic state through which the pulse drives the system.

The current-dependent cuts in Fig.~\ref{fig:switching_map}c resolve two families of behavior. Deep in the ordered phase, the response develops a sharp maximum near 35~mA and retains a large nonvolatile amplitude, consistent with switching between well-defined metastable configurations separated by comparatively large barriers (blue region). Closer to T$_N$, the maximum shifts to lower current and broadens while the written contrast is reduced, signaling a more weakly pinned and dynamically evolving state (red region). The extracted peak current I$_p$ and peak switching amplitude $(\Delta R_\perp/R_\parallel)_p$ in Fig.~\ref{fig:switching_map}d summarize this crossover: I$_p$ remains approximately constant in the low-temperature regime before decreasing rapidly near the transition, whereas the peak amplitude falls continuously on warming.

These observations define two dynamical writing regimes within the same ordered phase. At low temperature, the response is large and stable but requires a higher current scale, consistent with a relatively stiff magnetic background. Near T$_N$, smaller pulses can perturb the state, yet the written anisotropy is weaker and increasingly susceptible to relaxation. This distinction is central to the present work: the switching is controlled not only by the applied current, but by the temperature-dependent accessibility and stability of metastable magnetic configurations. The aging and post-field relaxation measurements presented next test this interpretation independently.

The evolution of the switching map suggests that the magnetic state changes from a strongly pinned low-temperature regime to a more weakly constrained and relaxation-prone state on approaching T$_N$. To determine whether this crossover reflects an intrinsic property of the magnetic landscape, rather than only the response of the microdevice to current pulses, we performed independent magnetization-aging and post-field-relaxation measurements on bulk crystals from the same parent growth.

Magnetic aging provides a direct test of history dependence. The crystal was zero-field cooled with a one-hour stop at a selected temperature T$_{\mathrm{pause}}$, after which cooling resumed to base temperature and the magnetization was measured on warming in a weak probe field. Subtracting an uninterrupted reference run isolates the memory retained from the thermal stop, $\Delta M(T)=M_{\mathrm{pause}}(T)-M_{\mathrm{reference}}(T)$ (Fig.~\ref{fig:afm_fm}a). A clear memory response persists throughout much of the ordered phase, demonstrating that the magnetic configuration does not equilibrate uniquely during cooling. Instead, the state reached below T$_N$ depends on its thermal history, as expected for a magnetic landscape containing multiple metastable configurations and a broad distribution of relaxation barriers.

The aging signal evolves in two characteristic forms (Fig.~\ref{fig:afm_fm}b). At lower pause temperatures, the response is dominated by a narrow memory dip centered near $T_{\mathrm{pause}}$, consistent with the retention of a thermally encoded state in a slowly relaxing magnetic landscape~\cite{Jonason1998MemoryChaosSpinGlasses, mydosh1993spin}. At higher pause temperatures, a broader step-like background becomes increasingly prominent, indicating that the waiting protocol modifies the magnetic response over a wider temperature interval. We therefore separate the aging signal phenomenologically into dip-like and step-like contributions, without assigning them to distinct microscopic phases. The relative weight of these two components changes across the same temperature range in which the switching map crosses over from a robust low-temperature response to the softer regime near T$_N$.

Post-field relaxation provides a complementary dynamical probe. After field cooling to a fixed temperature, the magnetic field was removed and the remanent magnetization was monitored for one hour (Fig.~\ref{fig:afm_fm}c). The continued evolution of $M(t)$ over minute timescales confirms that the ordered state remains dynamically active long after the external perturbation is removed~\cite{mydosh1993spin}. Rather than fitting the traces to a unique microscopic relaxation law~\cite{Bouchaud1996ModeCouplingAging}, we quantify their magnitude using a robust empirical amplitude, $A(T)$, evaluated over a common time window. The extraction procedure is described in the Relaxation protocol section of the Supplementary Information and in Supplementary Fig.~S15.

The connection to electrical switching becomes evident when $A(T)$ is compared with the inverse peak switching amplitude, $[(\Delta R_{\perp}/R_{\parallel})_p]^{-1}$ (Fig.~\ref{fig:afm_fm}d). The two quantities exhibit closely matching temperature dependencies throughout the ordered state: temperatures with stronger post-field relaxation correspond to a smaller electrically written resistance contrast, whereas the largest and most stable switching response occurs where the magnetic relaxation is weakest. This relationship does not imply that magnetization relaxation and electrical switching share a single microscopic timescale. Instead, it shows that both are governed by the same evolving magnetic landscape. Slow relaxation reduces the persistence of the electrically written anisotropy, while stronger freezing stabilizes the nonvolatile state.

Together, the aging, relaxation, and switching measurements establish a unified picture. Co$_x$TaS$_2$ near $x=0.33$ does not behave as a rigid AFM with a single current threshold. It hosts a history-dependent landscape whose dynamical character changes across the ordered phase. Deep below T$_N$, magnetic configurations are strongly pinned, producing a large and persistent written state once the relevant current scale is reached (blue region). Closer to T$_N$, the landscape becomes more accessible to weak perturbations but also more susceptible to relaxation, yielding lower activation currents together with a reduced nonvolatile contrast (red region). The current pulses therefore exploit, rather than simply overcome, the frustrated magnetic landscape.

The role of competing magnetic orders provides an important connection to the switching phenomenology of Fe$_x$NbS$_2$. In that system, the enhanced electrical response away from exact one-third intercalation was linked to a spin-glass component coexisting with long-range AFM order, whose collective dynamics participate in the switching process~\cite{EM,FeNbS2SpinGlassSciAdv}. Neutron-scattering measurements subsequently showed that small changes in Fe concentration strongly alter the balance between distinct magnetic structures, suggesting that competition between nearby ordered phases provides a microscopic source of the metastability and slow dynamics~\cite{wu2022highly}. These interpretations are complementary rather than mutually exclusive: phase competition and compositional disorder can generate a rugged energy landscape, while the glassy degrees of freedom provide the collective channel through which current pulses reorganize the magnetic state. Co$_x$TaS$_2$ realizes a distinct limit of this mechanism. Here, the switchable composition remains close to one-third intercalation but lies directly at the narrow boundary between Hall-active and Hall-inactive magnetic regimes. The enhanced switching therefore does not require moving far from commensurate filling; instead, it emerges where competing magnetic configurations are intrinsically closest in energy. Together with the observed aging and slow post-field relaxation, this comparison suggests that frustration at the composition-driven phase boundary plays a more direct role in assisting electrical reconfiguration in Co$_x$TaS$_2$.

\section{Conclusions}

We have demonstrated nonvolatile electrical switching in near-stoichiometric Co$_x$TaS$_2$, at a composition where the spontaneous Hall response is suppressed but electronic and magnetic signatures of nearby competing multi-Q states remain. Current pulses write anisotropic resistance states with an activation current density nearly an order of magnitude below that reported for switching in the lower-intercalation Hall-active regime~\cite{xiong2025all,zhang2026current}. The response is not described by a single threshold: instead, it separates into a robust low-temperature regime and a softer near-T$_N$ regime with reduced written contrast.

Independent magnetic measurements reveal the origin of this behavior. Thermal aging establishes memory and history dependence throughout the ordered phase, while post-field relaxation shows that the magnetic state continues to evolve over laboratory timescales. The close correspondence between relaxation strength and the inverse switching amplitude links the stability of the electrically written state to the same temperature-dependent magnetic landscape. These observations support frustration-assisted switching, in which current pulses access metastable magnetic configurations whose barriers and relaxation rates are controlled by proximity to competing orders.

This mechanism distinguishes the present response from electrical control of the Hall-active chiral phase through carrier-density tuning or Hall-state reversal~\cite{xiong2025all,zhang2026current, CoTaS2Gating3QNatCommun2025}. It also differs from a simple heat-assisted recording process, where transient heating suppresses magnetic order and an external field selects the final state~\cite{Rodriguez2026HAMR}. Local pulse heating may contribute to barrier crossing in Co$_x$TaS$_2$, but it does not alone account for the field-independent writing geometry, the AMR-like readout, or the correlated evolution of switching, aging, and relaxation. More broadly, the results show that magnetic frustration need not be detrimental to device functionality: a dynamically soft, history-dependent energy landscape can provide an additional pathway for electrically reconfiguring AFM states~\cite{FeNbS2SpinGlassSciAdv, haley2023long}.

\section*{Acknowledgements}

Work by M.S. and E.M. was performed with support from the European Research Council grant no. ERC-101117478, the Israeli Science Foundation grant no. ISF-885/23 and the PAZY foundation grant no. 412/23.
A portion of this work was performed at the National High Magnetic Field Laboratory, which is supported by National Science Foundation Cooperative Agreement No. DMR-2128556 and the State of Florida.

\section*{Conflict of Interest}

The authors declare no conflict of interest.

\section*{Data Availability Statement}

The data that support the findings of this study are available from the corresponding author upon reasonable request.

\section*{Supplementary Information}

Additional supplementary information can be found online in the Supplementary Information section. The Supplementary Information includes sample growth, structural characterization, and composition, focused-ion-beam microdevice fabrication and geometry, switching measurement and normalization protocols, current-density estimates, thermodynamic and ARPES methods, and aging and relaxation procedures.

\bibliographystyle{apsrev4-2}
\bibliography{ref}

\end{document}